\documentclass[fleqn,12pt,twoside]{article}
\usepackage[headings]{espcrc1}

\readRCS
$Id: espcrc1.tex,v 1.2 2004/02/24 11:22:11 spepping Exp $
\usepackage{graphicx}
\usepackage[figuresright]{rotating}

\newcommand{\AmS}{{\protect\the\textfont2
  A\kern-.1667em\lower.5ex\hbox{M}\kern-.125emS}}

\title{Anisotropic flow in AuAu and CuCu 
       at 62 GeV and 200 GeV}

\author{Gang Wang\address[MCSD]{Kent State University, Physics Department, Kent, OH 44242 USA}
        (for the STAR\thanks{For the full list of STAR authors and acknowledgements, see appendix `Collaborations' in this volume.} Collaboration)}
       
\runtitle{Anisotropic flow in AuAu and CuCu at 62 GeV and 200 GeV}
\runauthor{Gang Wang}

\begin{document}

% typeset front matter
\maketitle

\begin{abstract}
We present STAR's measurements of directed flow ($v_1$) and elliptic flow ($v_2$) for 
charged hadrons in AuAu collisions at 62 and 200 GeV, as a function of pseudorapidity, 
$p_t$ and centrality. $v_2$ results in CuCu collisions at 200 GeV are also presented. 
\end{abstract}

\section{MOTIVATION}
Directed flow and elliptic flow are quantified by the first harmonic ($v_1$) and the 
second harmonic ($v_2$), respectively, in the Fourier expansion of the azimuthal 
distribution of produced particles with respect to the reaction plane~\cite{Methods}, 
and carry information on the early stages of the collision~\cite{Whitepaper}. The 
shape of $v_1(y)$ is of special interest because it has been identified as a possible 
QGP signature~\cite{Whitepaper}. At much lower energies, $v_1(y)$ is an almost linear 
function of rapidity~\cite{v1LowerEnergies}, whereas at RHIC, it is predicted to have 
a more complex shape, with the region near midrapidity having a slope of smaller 
magnitude than elsewhere~\cite{wiggle,bleicher}. It may exhibit a characteristic 
wiggle~\cite{wiggle,bleicher,antiflow,third-component}, whereby directed flow changes 
sign at midrapidity and at two other rapidities other than the spectator regions.  At 
RHIC, the measurement and interpretation of elliptic flow has received much more 
attention than directed flow~\cite{Whitepaper}. When centrality is varied in a heavy 
system like AuAu, the resulting change in $v_2$ is caused by the variation of both 
the number of participants and their initial spatial eccentricity; by studying a 
lighter system, we can hope to gain a better understanding of the separate role of 
these two factors. In this presentation, we include an initial study of 
elliptic flow in 200 GeV CuCu collisions.  

\section{DATA SETS AND ANALYSES}
This study is based on five million 62 GeV AuAu events, three million 200 GeV AuAu 
events, and two million 200 GeV CuCu events, all from a minimum-bias trigger.  
All errors are statistical.  The centrality definition and cuts used are the same 
as in Ref.~\cite{Flow200GeV}.

The suppression of non-flow effects (azimuthal correlations unrelated to the reaction 
plane) is emphasized in this investigation.  $v_1$ results are obtained using the 
three-particle cumulant method ($v_1\{3\}$)~\cite{Borghini}, the event plane
method with mixed harmonics ($v_1\{\mathrm{EP}_1,\mathrm{EP}_2\}$)~\cite{Methods,Flow200GeV},
and for the first time at RHIC, the standard method~\cite{Methods} with the first-order
event plane reconstructed from spectator neutrons ($v_1$\{ZDC-SMD\})~\cite{ZDC}. 
$v_2$ results come from a diverse set of analyses: the standard method, the cumulant 
method ($v_2\{2\}$ and $v_2\{4\}$), $v_2$\{ZDC-SMD\}, the scalar product method 
($v_2$\{AuAu-pp\} and $v_2$\{CuCu-pp\})~\cite{Flow200GeV,qDist}, and a method that 
involves fitting the distribution of the lengths of the flow vectors normalized
by the square root of the multiplicity ($v_2$\{qDist\})~\cite{Flow200GeV,qDist}.

\section{RESULTS}

\begin{figure}[t]
\begin{minipage}[t]{78mm}
\includegraphics[width=1.0\textwidth]{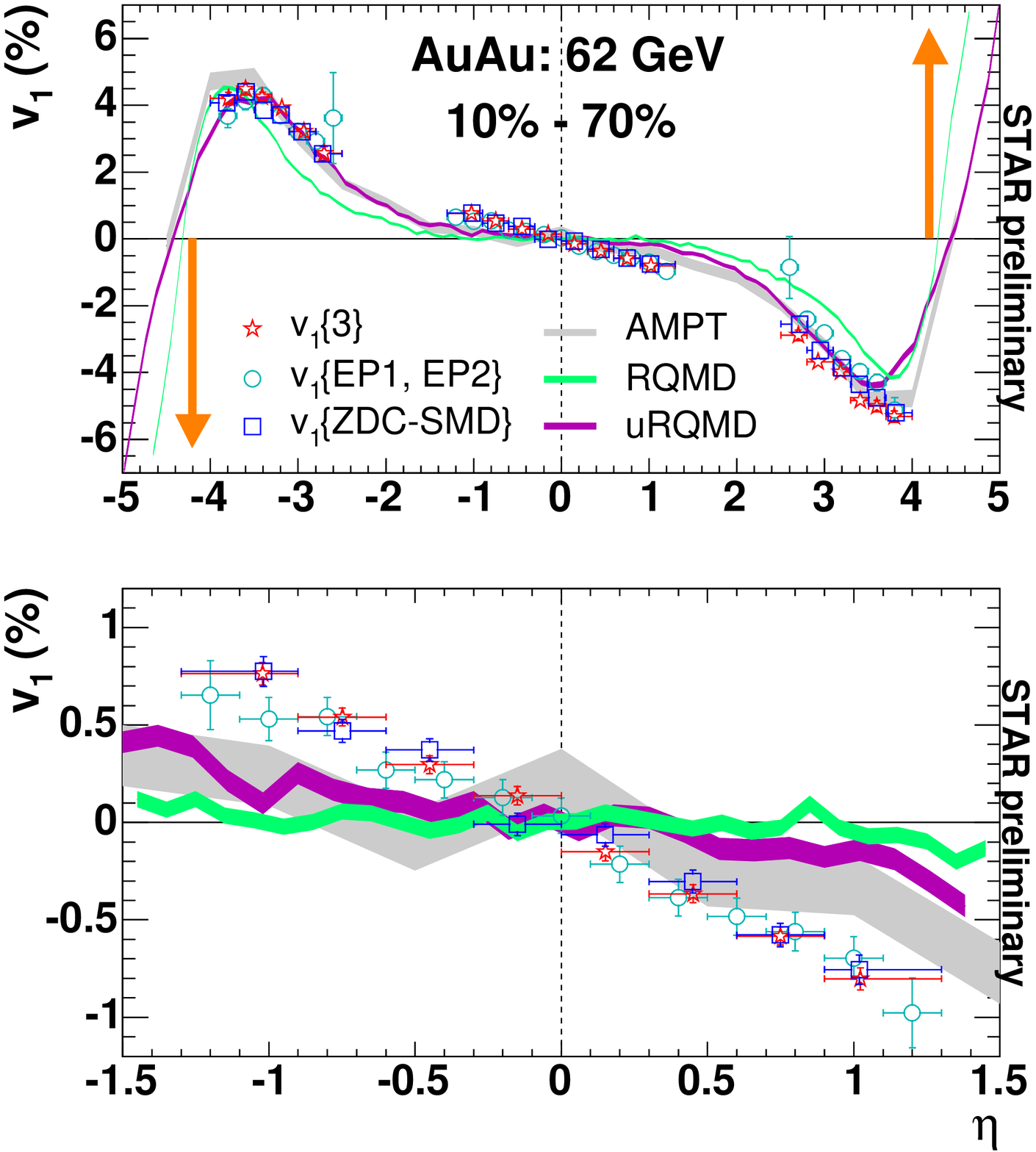}
\caption{Charged hadron $v_1$ vs. pseudorapidity, for 62 GeV AuAu. 
The lower panel shows the mid-pseudorapidity region in more detail.}
\label{fig:v1_62GeV}
\end{minipage}
\hfill
\begin{minipage}[t]{78mm}
\includegraphics[width=1.0\textwidth]{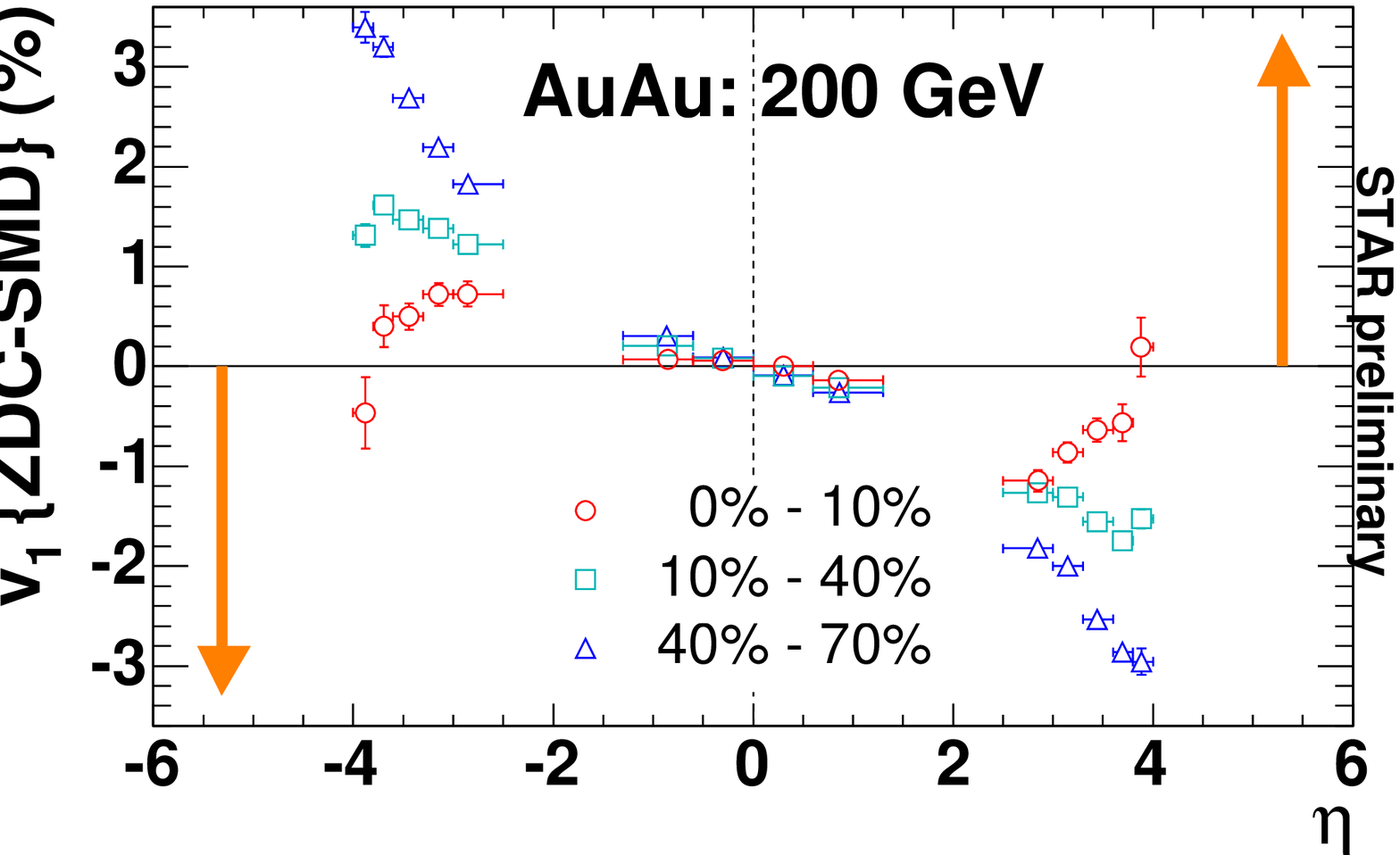}
\caption{Charged hadron $v_1$ vs. pseudorapidity, for 200 GeV AuAu. 
Here, only the $v_1$\{ZDC-SMD\} method is used, and three centrality 
intervals are plotted.}
\label{fig:v1_200GeV}
\end{minipage}
\vspace{-0.5cm}
\end{figure}

\subsection{Directed Flow}
\begin{figure}[t]
\begin{minipage}[t]{78mm}
\includegraphics[width=1.0\textwidth]{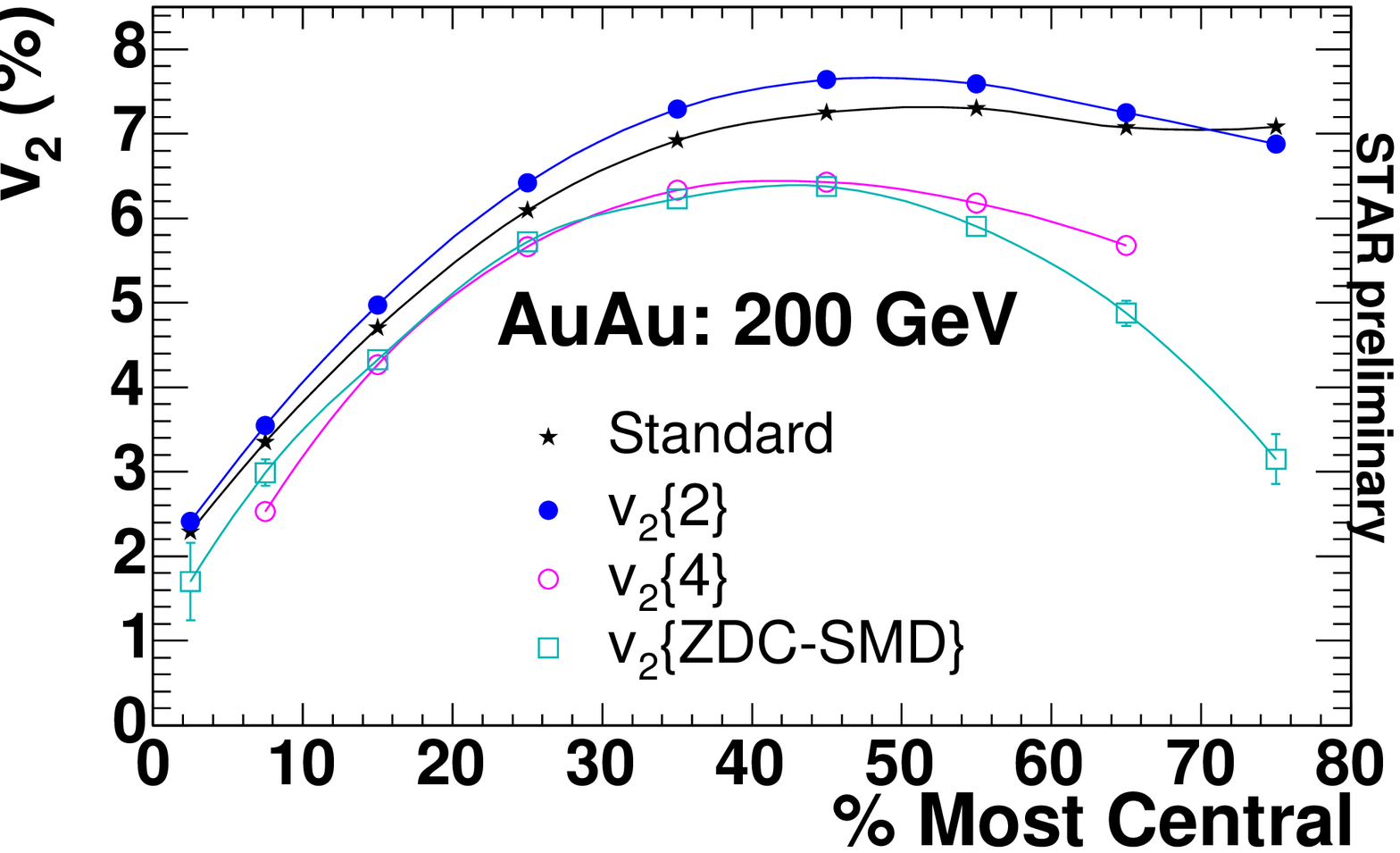}
\caption{Charged hadron $v_2$, integrated over $p_t$ and $\eta$, vs. centrality, 
for 200 GeV AuAu.} 
\label{fig:v2_int_1}
\end{minipage}
\hfill
\begin{minipage}[t]{78mm}
\includegraphics[width=1.0\textwidth]{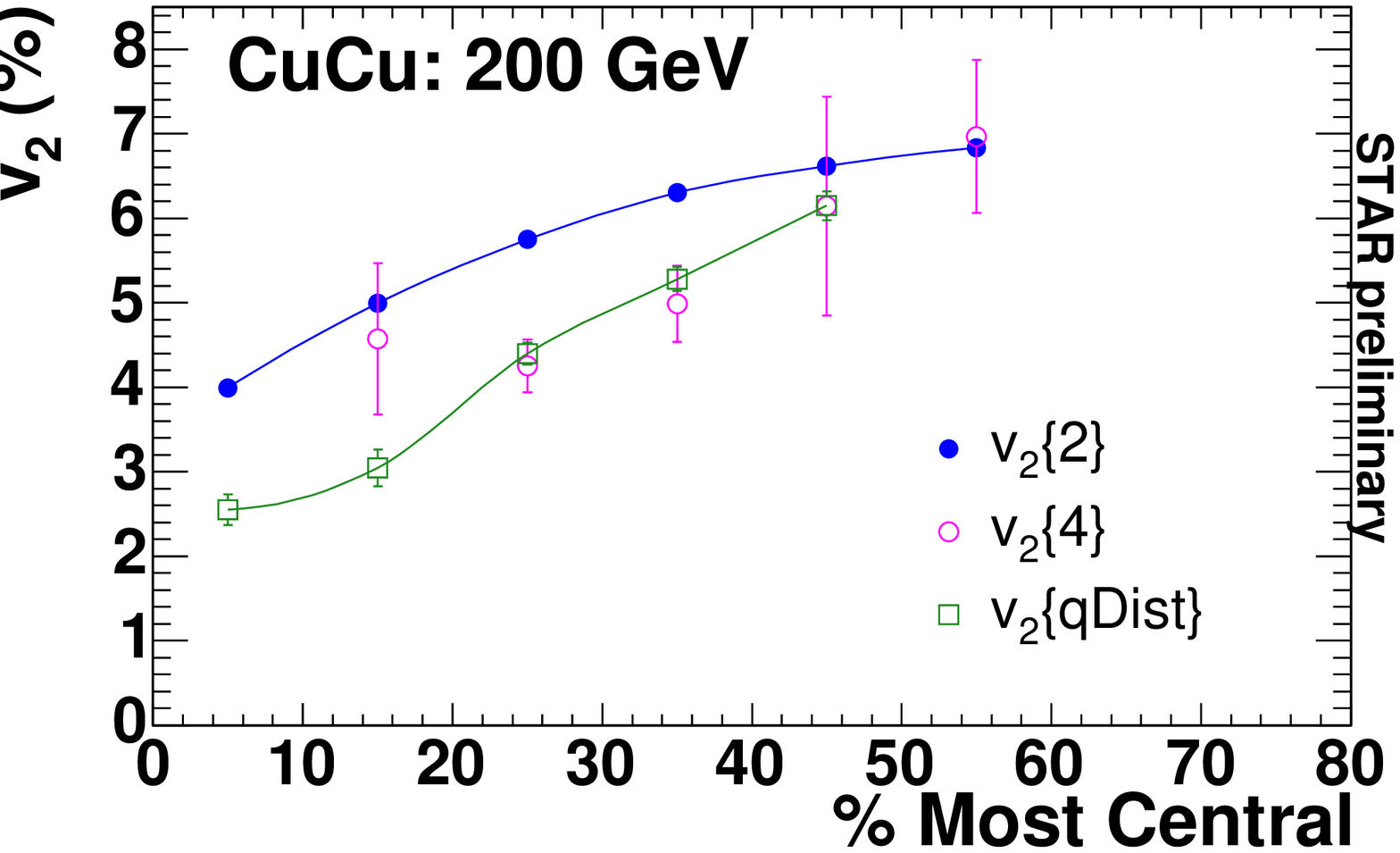}
\caption{Charged hadron $v_2$, integrated over $p_t$ and $\eta$, vs. centrality, 
for 200 GeV CuCu.} 
\label{fig:v2_int_2}
\end{minipage}
\end{figure}
Fig. \ref{fig:v1_62GeV} shows charged hadron $v_1$ as a
function of pseudorapidity, $\eta$, for centrality 10\%--70\% in 62 GeV AuAu collisions. 
The arrows indicate the direction of flow for spectators.
The arrow positions on the $\eta$ axis correspond to
where the incident ions lie on a rapidity scale.
STAR has a gap in $\eta$ coverage between the central Time Projection Chamber (TPC) and 
the Forward TPCs.  Results from the three different methods agree with each other very 
well.  AMPT~\cite{ampt}, RQMD~\cite{rqmd} and uRQMD~\cite{urqmd} model calculations 
all underpredict the charged hadron $v_1(\eta)$ within a unit or so of midrapidity, but 
then predict an increase in such a way that they come into good agreement with data 
over the region $2.5<|\eta|<4.0$.

Fig. \ref{fig:v1_200GeV} presents charged hadron $v_1$ vs. $\eta$ in three centrality 
bins for 200 GeV AuAu. The arrows have the same meaning as in Fig. \ref{fig:v1_62GeV}.
 The centrality dependence is qualitatively similar to the 
trend seen at SPS by NA49~\cite{NA49}.  STAR's coverage is such that we can probe the 
interesting region $|\eta| \sim 3.5$ to 4 where $v_1$ changes sign for centrality 0 
to 10\%. The sign change of $v_1$ may provide insights into the stronger 
stopping in more central collisions.

\subsection{Elliptic Flow}
Fig. \ref{fig:v2_int_1} presents charged hadron $v_2$, integrated over $p_t$ and 
$\eta$, vs. centrality for 200 GeV AuAu.  Results from the standard method, $v_2\{2\}$ 
and $v_2\{4\}$ are from RHIC Run II data reported in Ref.~\cite{Flow200GeV}.  A 
systematic check on $v_2$ measurements using the ZDC-SMD 1st harmonic event plane 
shows good agreement with $v_2\{4\}$, and reveals substantial non-flow effects in 
the standard method in peripheral collisions.

Charged hadron $v_2$ similarly integrated for 200 GeV CuCu is shown in Fig.~\ref{fig:v2_int_2}.
The methods $v_2\{4\}$ and $v_2$\{qDist\} should each have relatively little contribution 
from non-flow, and their agreement within errors is consistent with this assumption.  It 
is also evident that at 200 GeV, integrated elliptic flow is smaller in CuCu than in AuAu.  

\begin{figure}
\begin{minipage}[t]{78mm}
\includegraphics[width=1.0\textwidth]{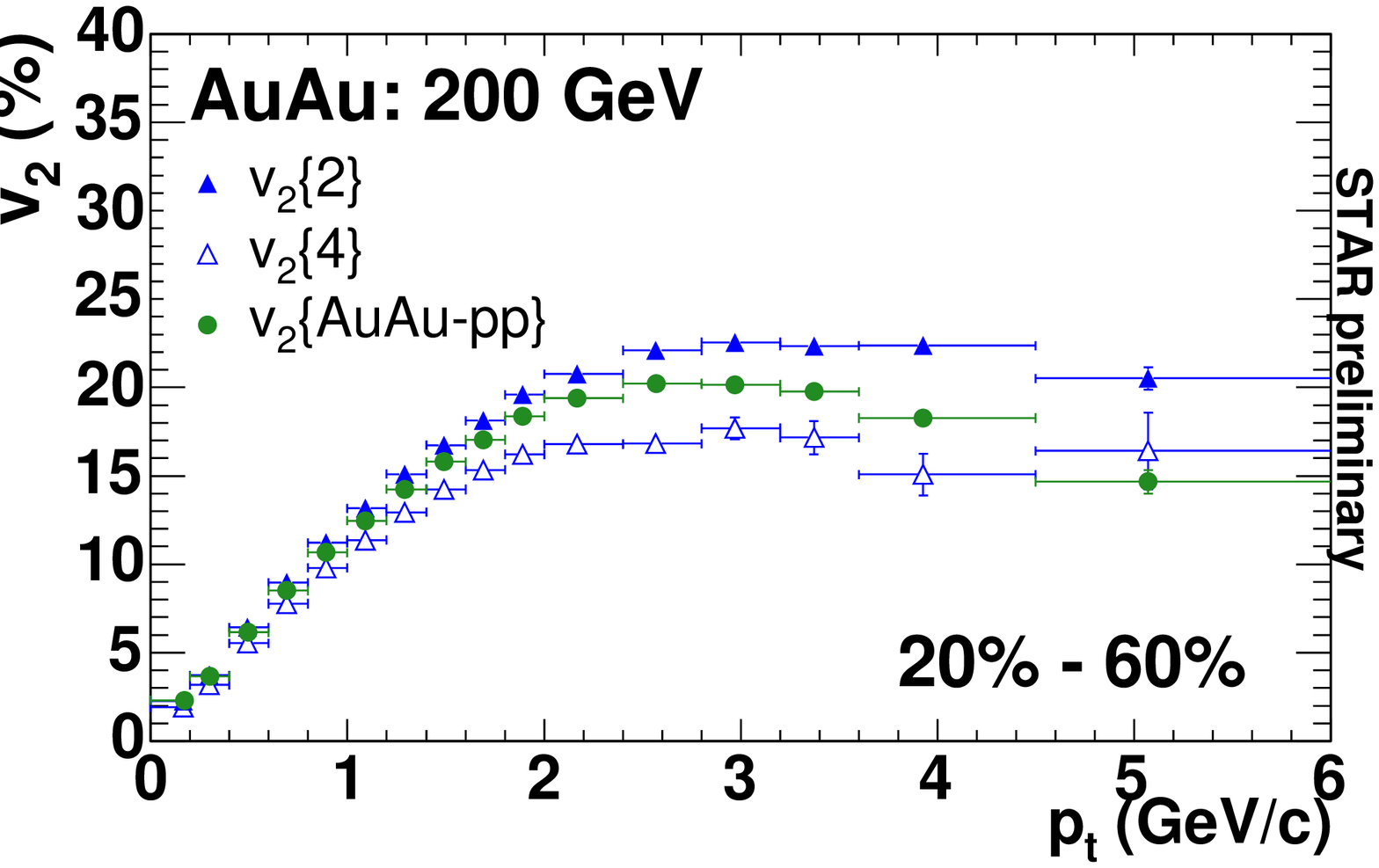}
\caption{Charged hadron $v_2$ vs. $p_t$, for 200 GeV AuAu.}
\label{fig:v2_pt_1}
\end{minipage}
\hfill
\begin{minipage}[t]{78mm}
\includegraphics[width=1.0\textwidth]{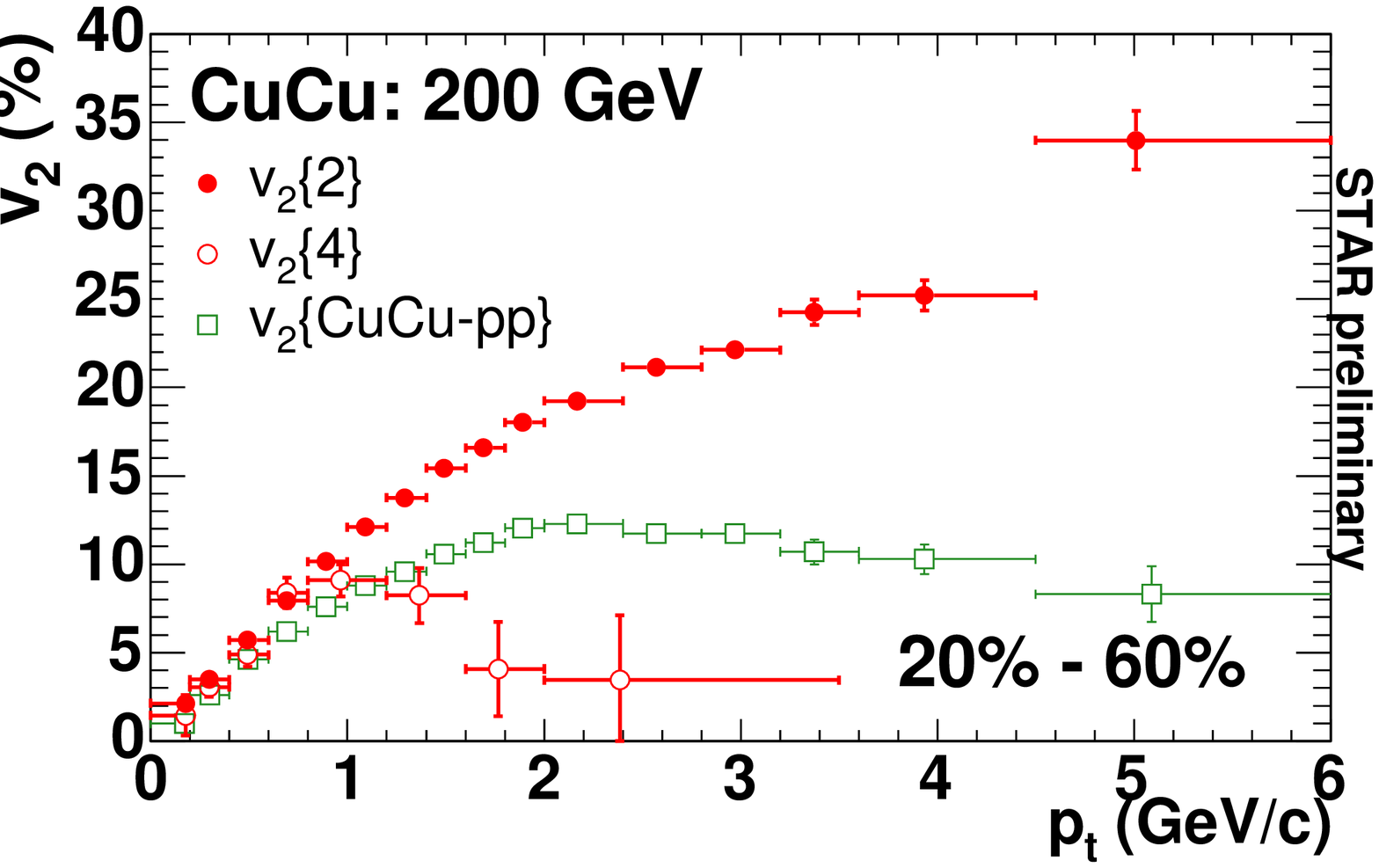}
\caption{Charged hadron $v_2$ vs. $p_t$, for 200 GeV CuCu.}
\label{fig:v2_pt_2}
\end{minipage}
\end{figure}

Fig. \ref{fig:v2_pt_1} and \ref{fig:v2_pt_2} show 200 GeV charged hadron $v_2$ vs. $p_t$, 
for AuAu and CuCu, respectively.  The difference between $v_2\{2\}$ and $v_2$\{CuCu-pp\} 
(believed to be mostly non-flow effects) increases with $p_t$ and becomes very large 
above about 1 GeV$/c$.  Comparing this pattern with AuAu, we conclude that the relative 
importance of non-flow is much larger in CuCu than in AuAu.  
In both systems, elliptic flow measurements based on $v_2\{4\}$ and $v_2$\{AA-pp\} demonstrate reasonable consistency 
across a region of $p_t$ (up to about 1.5 GeV$/c$).  However, the divergence between $v_2\{4\}$ and $v_2$\{CuCu-pp\} 
at higher $p_t$ requires further investigation.  Overall, at all $p_t$, $v_2\{4\}$ and 
$v_2$\{AA-pp\} are lower in CuCu than in AuAu.

%%We have presented STAR¡¯s measurements ofcharged hadron directed flow in AuAu collisions 
%%at 62 GeV and 200 GeV.
%%Elliptic flow results of charged hadrons in AuAu and CuCu at 200 GeV have also been presented.

\end{document}